\documentclass[prd, aps, nobibnotes, nofootinbib,
eqsecnum,
preprint,
showpacs,
preprintnumbers,
amsmath,
amssymb]{revtex4-1}

\usepackage{graphicx}
\usepackage{bm}
\usepackage{amssymb}
\usepackage[utf8]{inputenc}
\usepackage{hyperref}
\usepackage{mathrsfs}
\usepackage{xcolor}
\usepackage{subcaption}
\usepackage{multirow}
\usepackage{hhline}

\begin{document}

\title{{\bf{\Large Asymptotically safe cosmology with non-canonical scalar field}}}

\author{Rituparna Mandal}
\email{drimit.ritu@gmail.com}\,
\affiliation{School of Physics, University of Hyderabad, Central University P.O, Hyderabad 500046, Telangana, India}
\author{Soma Sanyal}
\email{somasanyal@uohyd.ac.in},\,
\affiliation{School of Physics, University of Hyderabad, Central University P.O, Hyderabad 500046, Telangana, India}


\begin{abstract}
We investigate the quantum modified cosmological dynamical equations in a Friedmann-Robertson-Walker universe filled with a barotropic fluid and a  general non-canonical scalar field characterized by a Lagrangian similar to k-essence model but with a potential term. Quantum corrections are incorporated by considering the running of gravitational and potential couplings, employing the functional renormalization group approach. Covariant conservation of the non-canonical scalar field and the background barotropic fluid is considered separately, imposing a constraint resulting from the Bianchi identity. This constraint determines the evolution of the cut-off scale with the scale factor and also reveals cosmic fixed points, depending on whether the flow ceases or continues to evolve. We explore how the general non-canonical scalar field parameter affects the different types of cosmic fixed points and how it differs from the canonical case. Furthermore, we establish a bound on the ratio of the RG parameters involving the non-canonical parameter for which the universe may exhibit accelerated expansion for mixed fixed points. This bound indicates the non-canonical scalar field includes larger sets of RG fixed point which may give rise to an accelerated universe.
\end{abstract}
\vskip 1cm
\maketitle

\section{Introduction}
A new era of cosmological study began with the observational discovery of the late-time cosmic acceleration reported in 1998, based on the type Ia supernovae (SN Ia) \cite{Riess:1998}. This has significantly transformed our understanding of the universe over the past years. The source responsible for this late-time acceleration, composed of gravitationally repulsive material, remains unclear and has been dubbed as the ``dark energy," constituting approximately 68$\%$ of the total energy density at present. The proposal of the cosmological constant, with a constant energy density, serves as a simple candidate for generating the late-time cosmic acceleration. This proposal is consistent with various observations, including the power spectrum of the CMB, the redshift of distant supernovae, and the distribution of the large-scale structure. However, the exceedingly small value of the cosmological constant, compared to the theoretical expectations without definitive reasons, has prompted theorists to explore a diverse array of alternative explanations. These alternatives include considering specific matter forms such as quintessence, k-essence, Chaplygin gas, Phantom in the matter sector, or modifications to General Relativity (GR) in the gravity sector.\\
In the matter sector, scalar field models play a significant role in cosmological models for understanding the accelerating universe, such as inflation in the early epoch \cite{Liddle:2000} and late-time acceleration by dark energy \cite{Tsujikawa:2006, Copeland:1998}. If the scalar field potential varies with energy scale or quantum effects are included through the coupling constant, the important question we want to ask is how these effects could impact the dark energy model of the scalar field.  
A useful and efficient way of including quantum effects is the Wilsonian Renormalization Group (RG) approach \cite{Wilson:1971}, which relies on coarse-graining procedures and produces the scale-dependent effective average action that can interpolate between short and long-distance scales. This effective average action includes all quantum fluctuations above a certain momentum scale, while quantum fluctuations below this scale are not included. Hence, this effective action describes the Renormalization Group trajectories, containing a number of coupling constants that run along with the cut-off scale, called the RG flow \cite{Reuter:1996cp, Souma:1999at, Lauscher:2001ya, Niedermaier:2006wt, Litim:2004}. In this approach, one can conceive of gravity free from divergence and predictive at a high energy scale, unlike the standard field-theoretical perturbative approach where gravity is non-renormalizable. The idea that gravity is non-perturbatively renormalizable and divergenceless at high energy, as the RG flow converges to a non-trivial UV fixed point, making it an ``Asymptotically Safe" quantum theory of gravity, was first proposed by S. Weinberg \cite{Weinberg:1976} in the late 1970s. \\
%
In the past decade, the scenario of asymptotically safe gravity, dictated by asymptotic safe fixed points at the UV scale, has led to the exciting idea that it may control the beginning of the universe \cite{Bonanno:2001xi}. This idea has been explored in the literature extensively, including those involving ideal fluid \cite{Bonanno:2001xi, Koch:2011}, even within the context of modified gravity theories like $f(R)$ theory. The Gaussian fixed point or infrared fixed point may provide the possibility of an accelerated universe, as observed today \cite{Weinberg:2010, Contillo:2011}. The functional RG technique in asymptotically safe gravity, where the cosmological constant and Newton's constant vary with the momentum scale, has been used to study deviations from standard cosmology \cite{Bonanno:2001xi, Shapiro:2005, Mandal:2020}. Moreover, quantum modifications to the cosmological evolution of the universe with a barotropic background fluid and the canonical, minimally coupled scalar field can yield the accelerated universe \cite{Hindmarsh:2011, Linder:2011}. In our case, we aim to investigate late-time acceleration by considering quantum gravitational effects within the framework of the exact renormalization group flow in a non-canonical scalar field model. \\
The motivation for introducing a non-canonical scalar field comes purely from phenomenological considerations, such as from the Born-Infeld action of string theory \cite{Sen:2002, Sachs:2003}, initially introduced in the context of inflation \cite{Mukhanov:1999, Mukhanov:1999a, Unnikrishnan:2012}. Later, it was extended for dark energy \cite{Chiba:2000, Steinhardt:2000, Steinhardt:2001, Chiba:2002, Scherrer:2004, Chimento:2004} models due to its intriguing characteristics as a modified matter model. Usually, the most general form for the Lagrangian of the non-canonical scalar field, containing non-canonical kinetic term and potential is in the form $\mathcal{L}(\phi, X) = f(\phi) F(X)-V(\phi)$ where $X=-\frac{1}{2}\partial_{\mu}\phi \partial^{\mu}\phi$ represents the kinetic energy, and $\phi$ is the scalar field. The Lagrangian without the need for a potential term ($V(\phi)=0$) can drive inflationary evolution, and it is referred to as k-essence \cite{Mukhanov:1999}. 
The Lagrangian, given by $\mathcal{L}(\phi, X) = F(X) - V(\phi)$ (with $f(\phi)=\text{constant}$), which we can view as a generalization of quintessence, plays a crucial role in studying dark energy models \cite{Wands:2013, Bose:2009}. It can also be employed to unify dark energy and dark matter models, as discussed in \cite{Cota:2011, Sahni:2017, Sahni:2021}. This model is referred to as the general non-canonical scalar field model, first studied in \cite{Fang:2007}. In this type of model, we have incorporated renormalization group corrections through Newton's constant and potential to investigate late-time effects in asymptotically safe gravity. We seek cosmic fixed points using a coupled dynamical equations system approach, examining how this cosmic fixed point could correspond to RG fixed points in the RG flow. Our focus is primarily on finding the cosmic fixed points that may lead to an accelerated universe in the infrared limit (corresponding to dark energy dominated universe) or the UV limit (inflating universe), depending on whether the cosmic fixed points correspond to the Gaussian RG fixed point or the UV fixed point. Our motivation lies in understanding how the general non-canonical scalar field parameters affect the cosmic fixed points and how this differs from the canonical case \cite{Hindmarsh:2011, Linder:2011}. \\
In this work, we shall employ the Einstein-Hilbert truncation \cite{Reuter:2001ag} in the gravity sector and incorporate corrections at the level of the Einstein equation, ensuring that the variation of the action with respect to the metric leaves the form of the Einstein field equations unaltered. The RG flow comes in the Einstein equations through the non-canonical scalar field model potential $V_{k}(\phi)$ and Newton's constant $G_{k}$. As here we have taken the covariant conservation of the background barotropic fluid and the non-canonical scalar field, it ensures that the Bianchi identity will provide us a constraint involving the RG parameters and the dimensionless cosmological variables (relative energy density of a component with respect to total energy density). This constraint is the same as the canonical scalar field case in \cite{Hindmarsh:2011} because the kinetic term in Lagrangian does not flow with the cut-off scale in the non-canonical scalar field similar to the canonical scalar field. Though the Bianchi identity poses the same constraint, we will see that the non-canonical parameter significantly changes the cosmic fixed point.  \\
In here, we incorporate the above ideas to formulate the cosmological evolution equations for dimensionless cosmological variables, which evolve with the scale factor for a universe that consists of a background barotropic fluid and the non-canonical scalar field. We find several cosmological fixed points from these cosmological dynamical equations using the formalism first described in \cite{Copeland:1998}, which may give an accelerated universe depending upon the diverse range of possible values for the RG parameters within the specific renormalization group theory. Moreover, we establish a  bound on the RG parameters involving the non-canonical parameter, which could confirm that the non-canonical scalar field model includes larger sets of RG fixed points that may give us an accelerated universe. \\   
The organization of this paper is as follows. In Sec.~\ref{sflrw}, we set up the Friedmann equations, the continuity equations, and the constraint equations between the RG parameters and dimensionless cosmological variables for a FLRW universe which consists of a barotropic fluid and the non-canonical scalar field taking into account quantum gravity corrections trough the scale-dependent Newton's constant $G_{k}$ and the potential of the non-canonical scalar field $V_{k}(\phi)$. In Sec.~\ref{dyeq}, we use these to write the quantum corrected cosmological evolution equations and find the different cosmic fixed points which may give an accelerated universe. We conclude with a discussion of our results. 
%
\section{RG improved cosmology}\label{sflrw}
Initially, we start by considering a spatially flat Friedmann-Lema\^itre-Robertson-Walker (FLRW) universe filled with background matter and a general non-canonical scalar field that interacts with gravity through the kinetic energy term. The Lagrangian for this non-canonical scalar field model is given by	
	\begin{align}
	\mathcal{L}(\phi, X)=F(X)-V_{k}(\phi)
	\label{lag}
	\end{align}
where $X=-\frac{1}{2}\partial_{\mu}\phi \partial^{\mu}\phi$ is the kinetic energy term.
We have taken Einstein-Hilbert truncation in the gravity sector where the quantum corrections are incorporated through the couplings including the gravitational couplings and the non-canonical scalar field potential amplitude, which run along with the momentum scale $k$. \\   
The general form of the field equations for the above-mentioned general action is given by 
\begin{align}
R_{\mu \nu}-\frac{1}{2}R g_{\mu \nu}=\frac{8 \pi G_{k}}{3}\left(T_{\mu \nu}^{\phi} + T_{\mu \nu}^{bm}\right)
\label{feq}
\end{align} 
where $T_{\mu \nu}^{\phi}$ describe the energy-momentum tensor for the non-canonical scalar field and $T_{\mu \nu}^{bm}$ is the background matter energy-momentum tensor in the form of a perfect fluid. Here, we have used the subscript $k$ on gravitational constant $G_{k}$ and potential $V_{k}$ to remind ourselves that these quantities flow with the momentum scale $k$.   
Note that the RG improvement has been made at the level of the field equations and is a kind of ``shortcut" that captures the effects of quantum gravity to the leading order.\\
For the non-canonical scalar field Lagrangian mentioned in \eqref{lag}, the energy-momentum tensor for this case is
\begin{align}
T_{\mu \nu}^{\phi} \equiv -\frac{2}{\sqrt{-g}}\frac{\delta \left(\sqrt{-g} \mathcal{L}\right )}{\delta g^{\mu \nu}}=g_{\mu \nu} \left[F(X)-V_{k}(\phi)\right]+ F_{,X} \partial_{\mu}\phi \partial_{\nu} \phi
\label{EM}
\end{align}
where ``${,X}$" denotes the partial derivative with respect to $X$. The energy-momentum tensor of a non-canonical scalar field is equivalent to a perfect fluid energy-momentum tensor  
\begin{align}
T_{\mu \nu}= (p+\rho)u_{\mu} u_{\nu}+p g_{\mu \nu}
\label{PE}
\end{align}
with velocity $u_{\mu}=\frac{\partial_{\mu}\phi}{\sqrt{2 X}}$, pressure $p_{\phi}(k)= \mathcal{L}=F(X)-V_{k}(\phi)$  and energy density
\begin{align}
\rho_{\phi} (k)=2X F_{,X} -p_{\phi} (k) = 2X F_{,X}-F(X)+V_{k}(\phi).
\label{enden}
\end{align}
In the FLRW universe, in the presence of a background matter field with energy density $\rho_{bm}$ and pressure $p_{bm}$, we obtain the modified Friedmann equations from \eqref{feq} as
	\begin{align}
		H^{2}&=\frac{8\pi G_{k}}{3} \left(2X F_{,X}-F(X)+V_{k}(\phi)+\rho_{bm}\right) \,  \label{flrw1st} \\ \dot{H}&=-4\pi G_{k} \left(2X F_{,X}+\rho_{bm}(1+\omega_{bm})\right) \,
		\label{flrw2nd}
	\end{align}
where $H=\frac{\dot{a}}{a}$ is the Hubble constant. \\
Another important point is to note that the left-hand side of the Einstein field equation is covariantly conserved, so by virtue of consistency, the right-hand side must also be covariantly conserved. This is called the Bianchi identity. Incorporating the Bianchi identity which implies  $\left[8 \pi G_{k} \left(T^{\phi \mu \nu}+T^{bm \mu \nu}\right) \right]_{;\nu}=0$, we obtain
\begin{align}
\frac{\partial \rho_{tot}(k)}{\partial t}=-3 H \rho_{tot}\left [ 1+\frac{p_{tot}(k)}{\rho_{tot}(k)}+\frac{1}{3}\frac{\partial\ln (G_{k} \rho_{tot})}{\partial \ln k} \frac{\mathrm{d} \ln k }{\mathrm{d} N}\right ]
\label{BI}
\end{align}
where $\rho_{tot}(k)=\rho_{\phi}(k)+\rho_{bm}$ and $N=\ln a$. We work here with the assumption that the energy-momentum tensor for the non-canonical scalar field and barotropic fluid is covariantly conserved separately, indicating no interaction between these two components. The covariant conservation of energy-momentum tensor for the non-canonical scalar field can be written as
\begin{align}
\frac{\partial \rho_{\phi}(k)}{\partial t}+3H\left (\rho_{\phi}(k)+p_{\phi}(k) \right )=0~,
\label{contem}
\end{align}
which in turn gives 
\begin{align}
\frac{\partial }{\partial N}\left [ 2 X F_{,X}-F(X)+V(\phi) \right ]+6F_{,X}=0~.
\label{conteX}
\end{align}
Further, the covariant conservation of the energy-momentum tensor of the background non-canonical scalar field yields
\begin{align}
\dot{\rho}_{bm}+3 H (p_{bm}+\rho_{bm})=0\,.
\label{matc}
\end{align} 
Since we have separately considered the energy-momentum tensor for the non-canonical scalar field and the barotropic fluid to be covariantly conserved, this leads to the consistency equation from the Bianchi identity as:
\begin{eqnarray}
 &\frac{\mathrm{d} \ln k }{\mathrm{d} N}  \frac{\partial\ln (G_{k} \rho_{tot})}{\partial \ln k} =0 \nonumber \\ \Rightarrow & \frac{\mathrm{d} \ln k }{\mathrm{d} N}  \left[ \frac{\mathrm{d}\ln G_{k} }{\mathrm{d} \ln k} + \frac{\kappa^{2} V_{k}}{3 H^{2}}\frac{\partial \ln V_{k} }{\partial \ln k}  \right] =0 .
\label{con}
\end{eqnarray}
Now, we introduce the dimensionless RG flow parameters arising due to the scale dependence through the gravitational constant and the non-canonical scalar field potential
\begin{align}
\eta_{RG}=\frac{\partial \ln G_{k}}{\partial \ln k},~~ \nu_{RG}= \frac{\partial \ln V_{k}}{\partial \ln k}~.
\label{RGP}
\end{align}
We find the consistency equation \eqref{con} in terms of RG flow parameters as
\begin{align}
\frac{\mathrm{d} \ln k }{\mathrm{d} N} \left[\eta_{RG}+ \frac{\kappa^{2} V_{k}(\phi)}{3 H^{2}} \nu_{RG} \right]=0~.
\label{flc}
\end{align}
This condition gives rise to two possibilities; one is $\frac{\mathrm{d} \ln k }{\mathrm{d} N}=0$ for all time, signifying the evolution of $\ln k$ with $N$. This comes to a halt at some freeze-in scale. Another possibility can be expressed as
\begin{align}
\eta_{RG}+ \frac{\kappa^{2} V_{k}(\phi)}{3 H^{2}} \nu_{RG}=0~.
\label{flowc}
\end{align}
These constraints on the RG flow parameters play an important role in understanding the cosmic dynamics or time evolution of the system when the quantum modifications are included through the running of the gravitational constant and the non-canonical scalar field potential.  
\section{Renormalization group improved system of dynamical equations} \label{dyeq}
In this section, we express the renormalization group improved system of dynamical equations to study the cosmic dynamics in the presence of a non-canonical scalar field and a background fluid. For this purpose, we introduce two dimensionless variables:  
\begin{align}
x=\frac{\kappa\sqrt{2 X F_{,X}-F}}{\sqrt{3} H}~, ~~~~ y=\frac{\kappa \sqrt{V_{k}(\phi)}}{\sqrt{3}H}~
\label{Ther}
\end{align}
where $\kappa^{2}=8 \pi G_{k} $.
Here, $x^{2}$ represents the relative energy density of the kinetic part of the scalar field $\rho_{ki}= 2 X F_{,X} -F$ defined in \eqref{enden}, and $y^{2}$ represents the relative energy density of the potential part defined as $\rho_{V}=V_{k}(\phi)$ with respect to the total energy density $\rho_{tot}(k)=\rho_{\phi}+\rho_{bm}=\frac{3 H^{2}}{\kappa^{2}}$ respectively. We define the equation of state for the kinetic part of the scalar field as:
\begin{align}
\omega_{ki}=\frac{F}{2 X F_{,X}-F}~.
\label{eoski}
\end{align}
Moreover, the equation of state of the non-canonical scalar field can be obtained in terms of the dimensionless variable $x$, $y$ and $\omega_{ki}$ as:
\begin{align}
\omega_{\phi}=\frac{p_{\phi}}{\rho_{\phi}}=\frac{\omega_{ki}x^{2}-y^{2}}{x^{2}+y^{2}}~.
\label{eos}
\end{align} 
Note that the $\omega_{\phi}$ has implicit scale dependence through dimensionless variables $x$ and $y$.
Now, differentiating the two dimensionless variables, namely, $x$ and $y$ with $N=\ln a ~(\mathrm{d}N=H\mathrm{d}t)$ , we obtain   
\begin{align}
\frac{\mathrm{d} x}{\mathrm{d} N} &= \frac{\kappa}{2 H}\frac{\left(2 X F_{,XX}+F_{,X} \right ) }{\sqrt{3 \left ( 2 X F_{,X}-F \right )}}\frac{\mathrm{d} X}{\mathrm{d} N}-x \frac{\dot{H}}{H^{2}} +\frac{x}{2}\frac{\partial \ln G_{k} }{\partial \ln k}\frac{\mathrm{d} \ln k }{\mathrm{d} N}   \label{evle1}\\ \frac{\mathrm{d} y}{\mathrm{d} N} &=\frac{y}{2}\left [ \frac{1}{V_{k}(\phi)}\frac{\partial V_{k}(\phi)}{\partial \phi}\frac{\dot{\phi}}{H} +\frac{\partial \ln V_{k}}{\partial \ln k} \frac{\mathrm{d} \ln k}{\mathrm{d} N} \right ] - y \frac{\dot{H}}{H^{2}}+ \frac{y}{2} \frac{\partial \ln G_{k}}{\partial \ln k} \frac{\mathrm{d} \ln k}{\mathrm{d} N}~. 
\label{evle2}
\end{align}
To substitute the factor $\frac{\dot{H}}{H^{2}}$ in the above evolution equations, one can obtain the factor $\frac{\dot{H}}{H^{2}}$ from \eqref{flrw2nd} by dividing $H^{2}$ in terms of the dimensionless variables, $x$ and $y$ as:
\begin{align}
\frac{\dot{H}}{H^{2}}=-\frac{3}{2} \left[(1+\omega_{bm})(1-y^{2})+x^{2}(\omega_{ki}-\omega_{bm})\right]~.
\label{Hsub}
\end{align}
To write the above equation, we have written the first Friedmann equation in terms of the dimensionless variable 
\begin{align}
x^{2}+y^{2}+\Omega_{bm}=1
\label{flrv}
\end{align}
where $\Omega_{bm}=\frac{\kappa^{2}\rho_{m}}{3 H^{2}}$ is the relative energy density of matter with respect to the total energy density.\\
Similar to the canonical scalar field \cite{Copeland:1998}, we define a variable that denotes the slope of the non-canonical scalar field potential, following \cite{Wands:2013}:
\begin{align}
\sigma=-\frac{1}{\kappa \sqrt{3 |\rho_{ki}|}} \frac{\partial V_{k}(\phi)}{\partial \phi} \frac{\sqrt{2 X}}{V_{k}(\phi)}~. 
\label{slopeV}
\end{align}
This variable is related as $\sigma=\sqrt{\frac{2}{3}}\lambda$ with the canonical scalar field variable $\lambda$ defined in \cite{Copeland:1998}, specifically when $\omega_{ki}=1$.\\
To express the first term of the evolution equations (\ref{evle1}, \ref{evle2}) in terms of the dimensionless variables $x$, $y$, $\sigma$, and the equation of state of the non-canonical scalar field, we derive two relations by rearranging Eq. \eqref{slopeV} and utilizing the continuity equation \eqref{conteX}:
\begin{align}
&\frac{\kappa V_{k,\phi} \sqrt{2X}}{2\sqrt{3V_{k}(\phi)}H^{2}}=-\frac{3}{2}\sigma x y \label{x1st}  \\ \frac{\mathrm{d} X}{\mathrm{d} N} =-&\frac{3F}{\left( 2 X F_{,XX}+F_{,X} \right ) \omega_{ki}}\left ( \omega_{ki}+1-\frac{\sigma y^{2}}{x} \right )
\label{y2nd}
\end{align}
so that the evolution equations \eqref{evle1} and \eqref{evle2} become 
\begin{align}
\frac{\mathrm{d} x}{\mathrm{d} N} &= \frac{3}{2}\left [ \sigma y^{2}-x (1+\omega_{ki})\right ] + \frac{3}{2} x \left[(1+\omega_{bm})(1-y^{2})+x^{2}(\omega_{ki}-\omega_{bm}) \right ] +\frac{x}{2} \eta_{RG}\frac{\mathrm{d} \ln k}{\mathrm{d} N} \label{xfinal} \\ \frac{\mathrm{d} y}{\mathrm{d} N} &= -\frac{3}{2} \sigma x y  + \frac{3}{2} y \left[(1+\omega_{bm})(1-y^{2})+x^{2}(\omega_{ki}-\omega_{bm}) \right ] +\frac{y (\eta_{RG}+\nu_{RG})}{2} \frac{\mathrm{d} \ln k}{\mathrm{d} N}~.
\label{yfinal}
\end{align} 
In this work, we assume that the renormalization group scale, to which the gravitational constant and non-canonical scalar field potential flow, is a function of cosmological time, denoted as $k=k(t)$, or alternatively expressed as $k=k(N)$. Here, we will discuss the evolution of the RG scale with time. To facilitate this, we rewrite Eq. \eqref{flowc} in terms of dimensionless variable $y$ as
\begin{align}
\eta_{RG}+y^{2} \nu_{RG}=0~.
\label{flrg}
\end{align}
For calculating the evolution of the RG scale with time, that is, $\frac{\mathrm{d} \ln k}{\mathrm{d} N}$ in terms of $x$, $y$ and $\sigma$, we differentiate Eq. \eqref{flrg} with respect to $N$ and obtain
\begin{align}
\frac{\mathrm{d}\ln k}{\mathrm{d}N}= \frac{1}{\alpha_{RG}}\left[\frac{3}{2}\frac{\epsilon_{RG}}{\nu_{RG}} \sigma x -\frac{3}{2}x^{2}(1+\omega_{ki})-\frac{3}{2}(1+\omega_{bm})\Omega_{bm} \right]
\label{fllnk}
\end{align}
where $\alpha_{RG}=\frac{1}{2}\left[\eta_{RG} + \nu_{RG} -\frac{\partial}{\partial \ln k}\ln \left(-\frac{\eta_{RG}}{\nu_{RG}}\right)\right]$. The full dynamical equations for $x$ and $y$ can be obtained by substituting \eqref{fllnk} into the right-hand sides of Eqs. (\ref{xfinal}, \ref{yfinal}). This evolution equation of $k$ with cosmic time is crucial because we must choose $k(N)$ in a manner that satisfies \eqref{fllnk} to maintain the consistency of the evolution equation. It's noteworthy that the non-canonical parameter plays a role in the evolution of the cut-off scale with cosmic time in Eq. \eqref{fllnk}, while the consistency condition \eqref{flrg} remains identical for both the canonical and non-canonical cases. \\
Another important point is to emphasize how to relate the RG scale to cosmological time or the Hubble parameter $H$. For this, one can show that $\frac{\partial \ln H}{\partial \ln k}=0$ when $\frac{\mathrm{d}\ln k}{\mathrm{d}N} \neq 0$, indicating that $H$ and $k$ are independent phase space variables. For this reason, the variation of the Hubble parameter with the scale factor changes exactly like the classical case:
\begin{align}
\frac{\mathrm{d} \ln H}{\mathrm{d} N} = -\frac{3(1+\omega_{ki})}{2}x^{2} -\frac{3}{2} (1+\omega_{bm}) \Omega_{bm}~.
\label{evh}
\end{align} 
However, the right-hand side of the above equation implicitly depends on the RG parameters. Comparing \eqref{evh} with \eqref{fllnk}, we indeed obtain: 
\begin{align}
\frac{\mathrm{d} \ln H}{\mathrm{d} N}= \alpha_{RG} \frac{ \mathrm{d} \ln k } { \mathrm{d} N} - \frac{3}{2} \frac{ \epsilon_{RG}}{ \nu_{RG}} \sigma x  
\label{flhlk}
\end{align}
which exactly matches with the canonical case. This relation for the non-canonical scalar field gives the condition that $\alpha_{RG}=1$ together with $\epsilon_{RG}=0$, $\frac{1}{\nu_{RG}}=0$, $x=0$ or $\sigma=0$ under which $k$ can be mapped with $H$. We will also see in the next section for which of these cosmic fixed points the RG scale can be chosen with the Hubble parameter. \\
Now, as we have not considered any form of $F(X)$ and $V_{k}(\phi)$, the evolution equations for $\omega_{ki}$ and $\sigma$ can be written as follows, 
\begin{align}
\frac{\mathrm{d} \omega_{ki}}{\mathrm{d} N}&=3\frac{2 \Sigma \omega_{ki} +\omega_{ki} -1 }{2\Sigma +1}\left ( \omega_{ki}+1-\frac{\sigma y^{2}}{x} \right )
\label{omega} \\
\frac{\mathrm{d} \sigma}{\mathrm{d} N}&=-3\sigma^{2}x(\Gamma-1)+\frac{3 \sigma \left (2 \Sigma (\omega_{ki}+1)+\omega_{ki}-1\right )}{2(2 \Sigma +1)(\omega_{ki}+1)}\left(\omega_{ki}+1-\frac{\sigma y^{2}}{x} \right ) \nonumber \\ & \qquad \qquad \qquad \qquad \qquad \qquad+ \sigma \left ( \epsilon_{RG}-\nu_{RG}-\frac{\eta_{RG}}{2} \right )\frac{\mathrm{d} \ln k}{\mathrm{d} N}~
\label{sigmaev}
\end{align}
where we have defined two new auxiliary variables in terms of the second derivative of $F(X)$ and potential $V_{k}(\phi)$ in such a way
\begin{align}
\Sigma \equiv \frac{X F_{,XX}}{F,X}~, \qquad
\Gamma \equiv \frac{V_{k} V_{k,\phi \phi}}{V_{k,\phi}^{2}}~.
\label{2ndG}
\end{align}
The above defined second order derivative variables will have evolution equations in terms of dimensionless variables $x$, $y$ and the new third order derivative variables. This procedure will repeat until we truncate this succession of equations by fixing the functions $F(X)$ and potential $V_{k}(\phi)$ of non-canonical Lagrangian terms. \\
For these, the first assumption is to consider the function that is taken to be 
\begin{align}
F(X)=A X^{n}
\label{F(X)ch}
\end{align}
where $A$ and $n$ are constants. For this choice, the equation of state for the kinetic part is $\omega_{ki}=\frac{1}{2n-1}$, and the evolution of the equation of state of the non-canonical scalar field $\omega_{ki}$ simply goes to zero, indicating $\frac{\mathrm{d}\omega_{ki}}{\mathrm{d}N}=0$. The evolution equation for the slope of the non-canonical scalar field $\sigma$ can be obtained as
\begin{align}
\frac{\mathrm{d}\sigma}{\mathrm{d}N}=-3\sigma^{2}x \left(\Gamma-1\right)+ \frac{3 \sigma (1-\omega_{ki})}{2(1+\omega_{ki})}\left(\omega_{ki}+1-\frac{\sigma y^{2}}{x}\right) + \sigma \left ( \epsilon_{RG}-\nu_{RG}-\frac{\eta_{RG}}{2} \right )\frac{\mathrm{d} \ln k}{\mathrm{d} N}~.
\label{chsigma}
\end{align}
To truncate the succession of the equations of the non-canonical scalar potential, the second assumption is that $\Gamma$ is some constant or exactly equal to one ($\Gamma=1$). If $\Gamma\neq 1$, but some other constant, the form of the potential can be taken as 
\begin{align}
V(\phi)=V_{0}\left(\phi -\phi_{0}\right)^{\frac{1}{1-\Gamma}}\,.
\label{V1}
\end{align}
For $\Gamma=1$, the potential form can be taken as an exponential potential 
\begin{align}
V(\phi)=V_{0}\exp(-\kappa \lambda \phi)\,.
\label{V2}
\end{align}
An essential point to consider is that, for a power-law potential where the Lagrangian is given by $\mathcal{L}=AX^{n}+V_{0} (\phi-\phi_{0})^{\delta}$ with $\delta=\frac{1}{(1-\Gamma)}$, the exponents of the kinetic and potential terms must satisfy the following relation from dimensional analysis \cite{Wands:2013}:
\begin{align}
\delta=\frac{n}{n+2}~.
\label{expodi}
\end{align} 
With this two choice of assumptions, we will see which form of potential is responsible for giving the stable attractor solution for the non-canonical scalar field dominated universe. 
\subsection{Cosmological Fixed Points }
In this subsection, we shall discuss the cosmic fixed points derived from the renormalization group improved dynamic equations involving the non-canonical scalar field and the background barotropic fluid. We are particularly interested in those cosmic fixed points which must be insensitive to initial conditions for describing the accelerated universe. We will also investigate whether the cosmic fixed points correspond to the RG fixed points at the infrared or UV scale.\\
As evident from \eqref{con}, two distinct types of cosmic fixed points emerge: one where the RG scale $\ln k$ becomes independent of $N$, and another where the RG scale maintains its scaling relationship with $N$ or cosmic time. \\
In the first case, where $\frac{\mathrm{d}\ln k}{\mathrm{d}N}=0$, the cosmological fixed points are the same as those calculated from the classical set of equations mentioned in \cite{Wands:2013}.  However, it is important to note that although the classical fixed point occurs when the flow of the RG scale with cosmic time stops at a certain energy scale, quantum modifications are still implicitly included in this system through the variables $x$, $y$, and $\sigma$. \\
A second type of cosmological fixed point can be calculated by taking the RG scale, which continues to evolve with the cosmological time as
\begin{align}
\frac{\mathrm{d}\ln k}{\mathrm{d}N}= \text{constant} \neq 0~.
\label{sfp}
\end{align}
Here, the dynamical equations of the dimensionless variables, namely $x$, $y$, and $\sigma$, explicitly depend on the cut-off scale $k$ through the RG parameters defined in \eqref{RGP}. Hence, to achieve a true cosmic fixed point, the RG parameters must cease to evolve with $k$ and become constant. This is only possible for RG fixed points, which may include Gaussian and non-Gaussian fixed points for the underlying quantum theory, where the coefficient of the right-hand side of Eq. \eqref{fllnk} simplifies to $\alpha_{RG}=\frac{1}{2}\left(\eta_{RG}+\nu_{RG}\right)$. \\ 
To proceed further, with the above-mentioned condition, we need to set (\ref{xfinal}, \ref{yfinal}), and (\ref{chsigma}) to zero to obtain the cosmic fixed points, corresponding to $x$, $y$, and $\sigma$ being constants. As the variables $x^{2}$ and $y^{2}$ are constrained by the equation in \eqref{flrv}, there can be three types of cosmic fixed points. Namely, when $x^{2}+y^{2}=1$, the total contribution of energy density comes from the non-canonical scalar field kinetic part and the potential energy density part. Another cosmic fixed point occurs when $x^{2}+y^{2}$ equals zero, signifying no contribution from the non-canonical scalar field. The last type of cosmic fixed point is when $x^{2}+y^{2}$ falls between zero and one, giving rise to scaling solutions where the energy density of the non-canonical scalar field scales at the same rate as the background fluid. \\
In the search for these cosmic fixed points, the trivial fixed point occurs when both $x_{*}=0$ and $y_{*}=0$ simultaneously, indicating no contribution from the non-canonical scalar field. In this scenario, the universe is background matter-dominated.\\
Now, if we seek the cosmic fixed point where the sole contribution comes from the non-canonical scalar field, characterized by $x^{2}+y^{2}=1$, three distinct situations arise. The first one corresponds to $x_{*}=0$ while $y_{*} \neq 0$. In this case, the RG modified dynamical equations for $x$ and $y$ take the form as
\begin{align}
\frac{\mathrm{d}x}{\mathrm{d}N}&=\frac{3}{2} \sigma y^{2} 
\label{pfixx} \\ \frac{\mathrm{d}y}{\mathrm{d}N}&= \frac{3}{2} y (1+\omega_{bm})(1-y^{2})+\frac{y}{2}(\eta_{RG}+ \nu_{RG} ) \frac{\mathrm{d}\ln k}{\mathrm{d} N}~.
\label{pfixy}
\end{align}
The above equations show that fixing $\sigma_{*}=0$ from the evolution equation of $x$ is necessary to attain a cosmic fixed point, given that $y_{*} \neq 0$. Additionally, to achieve $\frac{\mathrm{d}y}{\mathrm{d}N}=0$, we must fix $y_{*}=1$ or $\omega_{bm}=-1$, along with the condition $\eta_{RG}=-\nu_{RG}$. \\
An important point to note is that the consistency equation \eqref{flowc} always holds $y^{2}=-\frac{\eta_{RG}}{\nu_{RG}}$, restricting only one fixed point as $y_{*}=1$, as the condition $\eta_{RG}=-\nu_{RG}$ must be satisfied consistently. This is in contrast to the classical case mentioned in \cite{Wands:2013}, where another fixed point exists for this scenario with $y$ being arbitrary and $\omega_{bm}=-1$. At this potential-dominated fixed point where $x_{*}=0$ and $y_{*}=1$, the equation of state for the non-canonical scalar field becomes $\omega_{\phi}=-1$, potentially leading to an accelerated universe. To confirm this result, it is necessary to express the effective equation of state, denoted as $\omega_{eff}$, in the form
\begin{equation}
\omega_{eff}=\frac{p_{tot}}{\rho_{tot}}=\frac{p_{\phi}+p_{bm}}{\rho_{\phi}+\rho_{bm}}=\omega_{\phi} \Omega_{\phi}+\omega_{bm} \Omega_{bm}
\label{eoeff}
\end{equation}
where $\Omega_{\phi}=x^{2}+y^{2}$ and the equation of state of the background fluid is denoted as $\omega_{bm}=\frac{p_{bm}}{\rho_{bm}}$. In this case, with $\omega_{eff}=-1$, it becomes intriguing from a cosmological perspective as it opens up the possibility to describe dark energy or inflation phenomena depending on the different possible values of RG parameters for the specific renormalization group theory. \\
In the second case, where $y_{*}=0$ and $x_{*} \neq 0$, the total energy density solely comes from the kinetic part of the non-canonical scalar field. The dynamical equations for $x$ and $y$ evolve as
\begin{align}
\frac{\mathrm{d}x}{\mathrm{d}N} &=\frac{3x}{2} \left(\omega_{ki}-\omega_{bm} \right) \left(x^{2}-1 \right)+\frac{x}{2} \eta_{RG} \frac{\mathrm{d}\ln k}{\mathrm{d} N} 
\label{kfixx} \\ \frac{\mathrm{d}y}{\mathrm{d}N}&= 0~.
\label{kfixy}
\end{align}
Now, $\frac{\mathrm{d}x}{\mathrm{d}N}$ will be zero for two cases: either with $x_{*}=1$ or $\omega_{ki}=\omega_{bm}$ along with the condition $\eta_{RG} \rightarrow 0$. The condition $\eta_{RG} \rightarrow 0$ is crucial, as for a kinetic-dominated universe, these cosmic fixed points coincide with the RG fixed point that arises when gravity is classical, where no renormalization group improvement is taken into account. However, it is important to remember that although the fixed point matches the classical case, the RG modification is included in the theory. These two kinetic-dominated fixed points are achievable near infrared, where the gravitational coupling hits the Gaussian RG fixed point and becomes constant. The first cosmic fixed point occurs when $x_{*}=1$, $y_{*}=0$, along with $\eta_{RG} \rightarrow 0$, and it is completely dominated by the kinetic energy density of the non-canonical scalar field. Here, the effective equation of state is $\omega_{eff}=\omega_{ki}$, calculated from \eqref{eoeff}, since the equation of state of the non-canonical scalar field is $\omega_{\phi}=\omega_{ki}$. It is important to note that this fixed point does not result in an accelerated universe, as for an accelerated universe, $\omega_{eff}<-\frac{1}{3}$, which, in turn, implies $n<-1$ in \eqref{F(X)ch}. In this case, the cut-off scale evolves with time according to the following expression:
\begin{align}
\frac{\mathrm{d}\ln k}{\mathrm{d}N}=\frac{3}{\nu_{RG}}\left[\frac{\epsilon_{RG}}{\nu_{RG}}\sigma_{*}-(1+\omega_{ki})\right]
\label{klnk}
\end{align} 
where $\sigma_{*}$ can be fixed setting $\frac{\mathrm{d}\sigma}{\mathrm{d}N}=0$ in Eq. \eqref{chsigma}. Solving for $\sigma_{*}$ from $\frac{\mathrm{d} \sigma}{\mathrm{d}N}=0$, one can obtain either $\sigma_{*}=0$, or $\sigma_{*}$ can be fixed through the equation:
\begin{align}
-3 \sigma_{*} \left(\Gamma -1 \right)+\frac{3}{2}\left(1-\omega_{ki}\right)+3\left(\frac{\epsilon_{RG}}{\nu_{RG}}-1 \right) \left[\frac{\epsilon_{RG}}{\nu_{RG}}\sigma_{*} -\left(1+\omega_{ki}\right)\right]=0~.
\label{kisigma}
\end{align} 
Now, solving for $\sigma_{*}$ from this equation, its value can be calculated depending on the potential form. For example, for an exponential potential with $\Gamma=1$, $\sigma_{*}$ can be expressed in terms of the non-canonical scalar field parameter and RG parameters as: $\sigma_{*}=\frac{2\left(\frac{\epsilon_{RG}}{\nu_{RG}}-1\right)(1+\omega_{ki})-(1-\omega_{ki})}{2\left(\frac{\epsilon_{RG}}{\nu_{RG}}-1\right)\frac{\epsilon_{RG}}{\nu_{RG}}}$. Similarly, $\sigma_{*}$ can be determined for a power-law potential where $\Gamma$ is a constant, expressed in terms of $\omega_{ki}$ and RG parameters. Consequently, we can conclude that the kinetic-dominated fixed point can be achieved for both exponential and power-law potentials with different values of $\sigma_{*}$.\\
In the second cosmic fixed point, where $x$ is arbitrary with $\omega_{\phi}=\omega_{bm}$ and $y_{*}=0$, along with $\eta_{RG} \rightarrow 0$, the equation of state for background matter is exactly equal to the equation of state for the non-canonical scalar field. This type of solution has been referred to as a kinetically driven scaling solution, as mentioned in \cite{Wands:2013}. Note that for this fixed point, the arbitrariness of $x$ can be constrained by the first Friedmann equation $x_{*}^{2}+\Omega_{bm}=1$. \\
%
\begin{table}[h!]
\centering
 \begin{tabular} {|p{.6cm} |p{2.3cm} | p{1.5cm} | p{.5cm} | p{2.8 cm} |p{3cm} | p{3cm} | p{1.5cm} | p{1.5cm}| }
 \hline 
 case & $x^{2}$ & $y^{2}$ & $\sigma$ & $\Omega_{bm}$ & $\omega_{eff}$ & $\frac{\mathrm{d}\ln k}{\mathrm{d}N}$ & existence & type \\ [0.7 ex] 
 \hline
a & 0 & 1 & 0 & 0 & -1 &  & {\tiny $\frac{\eta_{RG}}{\nu_{RG}}=-1$ } & Potential \\ 
 b1 & 1 & 0 & $\sigma_{*}$ & 0 & $\omega_{ki}$ & {\tiny $ \frac{3}{\nu_{RG}}\left[\frac{\epsilon_{RG}}{\nu_{RG}}\sigma_{*}-(1+\omega_{ki})\right] $} & $\eta_{RG} \rightarrow 0 $  &  Kinetic \\
b2 & arbitrary & 0 & 0 & $1-x_{*}^{2}$ & $\omega_{bm}$ & $ -\frac{3}{\nu_{RG}} \left(1+\omega_{bm} \right) $ & $\eta_{RG} \rightarrow 0$ &   {\tiny Kinetic Scaling }\\
c1 & $ 1+ \frac{\eta_{RG}}{\nu_{RG}}$ & $-\frac{\eta_{RG}}{\nu_{RG}}$ & 0 & 0 & $ \omega_{ki}+\frac{\eta_{RG}}{\nu_{RG}}(1+\omega_{ki})$ & $-\frac{3\left(1+\omega_{ki}\right)}{\nu_{RG}}$ &  $\Gamma (0)=0$ &  Mixed \\
c2 & $ 1+ \frac{\eta_{RG}}{\nu_{RG}}$  & $-\frac{\eta_{RG}}{\nu_{RG}}$ & $\sigma_{*}$ & 0 & $ \omega_{ki}+\frac{\eta_{RG}}{\nu_{RG}}(1+\omega_{ki})$ & { \tiny $\frac{3}{\nu_{RG}} \left[\frac{\sigma_{*}}{x_{*}}- \left(1+\omega_{ki}\right) \right]$ } & {\tiny $\epsilon_{RG}=\nu_{RG}$ } & Mixed  \\ 
d1 & 0 & $-\frac{\eta_{RG}}{\nu_{RG}}$ & 0 & { \tiny $ 1+ \frac{\eta_{RG}}{\nu_{RG}}$ } & { \tiny $\omega_{bm}+\frac{\eta_{RG}}{\nu_{RG}}(1+\omega_{bm})$ } & $-\frac{3\left(1+\omega_{bm}\right)}{\nu_{RG}}$ & $\omega_{bm} \neq 0$ & Scaling \\
d2 & { \tiny $ \frac{\left(1+\omega_{bm}\right)}{\left(\omega_{bm}-\omega_{ki}\right)}\left(\frac{\eta_{RG}}{\nu_{RG}}\right) $ } & $-\frac{\eta_{RG}}{\nu_{RG}}$ & $\sigma_{*}$ & { \tiny $1 - \left(\frac{\eta_{RG}}{\nu_{RG}}\right) \frac{\left(1+\omega_{ki}\right)}{\left(\omega_{bm}-\omega_{ki}\right)}$ } & $\omega_{bm}$ & 0 & {\tiny $ \epsilon_{RG}= \nu_{RG}$ } & Scaling  \\[1ex] 
 \hline
\end{tabular}
\caption{Renormalization group improved cosmological fixed points for Einstein gravity with non-canonical scalar field and background barotropic fluid.}
\label{table-I} 
\end{table}
Another important type of cosmic fixed point is the mixed fixed point, where the total energy density comes from both the kinetic and potential energy density, satisfying $\Omega_{\phi}=x^{2}+y^{2}=1$, but with no barotropic fluid components ($\Omega_{bm}=0$). For this kind of cosmic fixed point, $y$ can always be determined from the consistency condition \eqref{flrg} as $y_{*}=\pm \left(-\frac{\eta_{RG}}{\nu_{RG}}\right)^{\frac{1}{2}}$, which in turn, fixes $x_{*}=\left(1+\frac{\eta_{RG}}{\nu_{RG}}\right)^{\frac{1}{2}}$. Here, for these value of $x_{*}$ and $y_{*}$, the evolution equations of $x$ and $y$ reduce to
\begin{align}
\frac{\mathrm{d}x}{\mathrm{d}N} &=\frac{3}{2} \sigma x^{2} y^{2} \nu_{RG} \left(1- \frac{\epsilon_{RG}}{\nu_{RG}}\right)
\label{mfixx} \\ \frac{\mathrm{d}y}{\mathrm{d}N}&= -\frac{3}{2}\sigma x y (1-\frac{\epsilon_{RG}}{\nu_{RG}})~.
\label{mfixy}
\end{align}
To obtain a cosmic fixed point, we need to set $\frac{\mathrm{d}x}{\mathrm{d}N}=0$ and $\frac{\mathrm{d}y}{\mathrm{d}N}=0$, which yield two fixed points. One is for $\sigma_{*}=0$, and the other is for the condition $\epsilon_{RG}=\nu_{RG}$. At one fixed point where $\sigma_{*}=0$, the RG scale parameter evolves as:
\begin{align}
\frac{\mathrm{d}\ln k}{\mathrm{d}N}=-\frac{3\left(1+\omega_{ki}\right)}{\nu_{RG}}~.
\label{lnkm}
\end{align}
For $\omega_{ki}=1$, the evolution of the RG scale matches with the result of the canonical scalar field mentioned in \cite{Hindmarsh:2011}. For another fixed point where $\epsilon_{RG}=\nu_{RG}$, $\sigma_{*}$ can be derived for the above-mentioned $x_{*}$ and $y_{*}$ from the equation:
\begin{align}
-3\sigma_{*}x_{*}(\Gamma-1)+\frac{3(1-\omega_{ki})}{2(1+\omega_{ki})}\left(1+\omega_{ki}-\sigma_{*} \frac{y_{*}^{2}}{x_{*}}\right)-\frac{\eta_{RG}}{2}\frac{\mathrm{d}\ln k}{\mathrm{d}N}=0
\label{sig2m}
\end{align}
where the evolution of the cut-off scale can be written for this fixed point as
\begin{align}
\frac{\mathrm{d}\ln k}{\mathrm{d}N}=\frac{1}{\alpha_{RG}}\left[\frac{3}{2}\sigma_{*}x_{*}-\frac{3}{2}x_{*}^{2}(1+\omega_{ki}) \right]~.
\end{align}
Similar to the kinetically-driven fixed point, the value of $\sigma_{*}$ can be derived from the above equations, depending on the potential profile. This dependency is influenced by both the exponential potential ($\Gamma=1$) and the power-law potential ($\Gamma=\text{constant}$).\\
For these two fixed points, the important result comes from the effective equation of state, which can be written from \eqref{eoeff} using \eqref{eos} as 
\begin{align}
\omega_{eff}=\omega_{ki}+\frac{\eta_{RG}}{\nu_{RG}}(1+\omega_{ki})~.
\label{effeos}
\end{align}
These cosmic fixed points may offer a dark energy solution with a variety of equations of state for different values of $\omega_{ki}$ and various renormalization group theories, where the RG parameters could assume different possible values. Moreover, it is very interesting to note that the condition which gives rise to cosmic acceleration must satisfy $\omega_{eff}<-\frac{1}{3}$, which in turn gives a bound on  $\frac{\eta_{RG}}{\nu_{RG}}$ in terms of the non-canonical scalar field parameter $\omega_{ki}$ 
\begin{align}
\frac{\eta_{RG}}{\nu_{RG}} < -\frac{\left(1+ 3\omega_{ki} \right)}{3 \left(1+\omega_{ki}\right)}~.
\label{bndrg} 
\end{align}
For a canonical scalar field, the bound on the RG parameter is $\frac{\eta_{RG}}{\nu_{RG}} < -\frac{2}{3}$ for which the expansion of the universe is accelerating. Here, this bound is particularly important because it gives a general bound of RG parameters in terms of non-canonical scalar field kinetic term parameter $n$ or $\omega_{ki}$ of Eq. \eqref{F(X)ch}. \\
Another interesting point to note is that at the asymptotically safe matter fixed point \cite{Percacci:2003, Narain:2010, Rahmede:2010} where all the couplings of the scalar field vanish at the UV scale, the RG parameters assume values such as $\eta_{RG}\rightarrow -2$ and $\nu_{RG} \rightarrow 4$. This scenario, however, does not result in an accelerated universe, as $\omega_{eff}=0$ for the canonical case where $\omega_{ki}=1$. Surprisingly, for the non-canonical scalar field, it is observed that for the same RG parameters, an accelerating universe can be achieved at the UV scale for $\omega_{ki}<\frac{1}{3}$. This, in turn, imposes a bound on the exponent of the non-canonical kinetic term in Eq. \eqref{F(X)ch} as $n>2$. \\
Next, we come to scaling fixed points where the non-canonical scalar field energy density follows the energy density of background matter. This may be radiation or matter. As usual, the consistency condition \eqref{flrg} makes sure that $y_{*}=\left(-\frac{\eta_{RG}}{\nu_{RG}}\right)^{\frac{1}{2}}$. From Eq. \eqref{flrv}, we can fix $x_{*}=0$ and $\Omega_{bm}=1-y_{*}^{2}=1+\frac{\eta_{RG}}{\nu_{RG}}$. For this particular scaling fixed point, the evolution equations for $x$ and $y$ take the form   
\begin{align}
\frac{\mathrm{d}x}{\mathrm{d}N} &=\frac{3 }{2} \sigma y^{2}
\label{s1fixx} \\ \frac{\mathrm{d}y}{\mathrm{d}N}& = 0~.
\label{s1fixy}
\end{align}
From the above equations, we can easily see that $\sigma_{*}=0$ for getting $\frac{\mathrm{d}x}{\mathrm{d}N}=0$. Here, the effective equation of state $\omega_{eff}$ can be derived from Eq. \eqref{effeos} as
\begin{align}
\omega_{eff}=\omega_{bm}+\frac{\eta_{RG}}{\nu_{RG}}\left(1+\omega_{bm}\right)~.
\label{scaeff}
\end{align}
From here, we can see that unlike the mixed fixed point, the effective equation of state does not depend on $\omega_{ki}$. Instead, it depends on the equation of state of the background matter. However, in this case, the condition on $\frac{\eta_{RG}}{\nu_{RG}}$ for the accelerating universe can be set as Eq. \eqref{bndrg}, the only change being that $\omega_{ki}$ is replaced by $\omega_{bm}$. Here, at this fixed point, the RG scale evolution is fixed as  
\begin{align}
\frac{\mathrm{d}\ln k}{\mathrm{d}N}=-\frac{3\left(1+\omega_{bm}\right)}{\nu_{RG}}~.
\label{56} 
\end{align}
It is important to note here that at the asymptotic matter fixed point in the UV scale where $\eta_{RG}=-2$ and $\nu_{RG}=4 $ and $\alpha_{RG}=1$, we obtain the scaling relation $k \propto H $ from Eq. \eqref{flhlk} as $x_{*}=0$ for this fixed point.  This scaling cosmic fixed point is exactly the same as the canonical case and can be mapped with the asymptotically safe matter fixed point in the UV scale \cite{Hindmarsh:2011}. \\
Another general scaling fixed point can be written by setting $\omega_{\phi}=\omega_{bm}$ in Eq. \eqref{eos} which in turn gives
\begin{align}
x_{*}^{2}=\frac{\left(1+\omega_{bm}\right)}{\left(\omega_{bm}-\omega_{ki}\right)}\left(\frac{\eta_{RG}}{\nu_{RG}}\right)~.
\label{scax}
\end{align}
Obviously, to get the $x_{*}$, we have used $y_{*}^{2}=-\left(\frac{\eta_{RG}}{\nu_{RG}}\right)$ from consistency condition. The value of $x_{*}$ and $y_{*}$ will definitely set the fixed point value of the density parameter of $\Omega_{bm}$ in terms of RG parameter, non-canonical scalar field parameter and the background equation of state as
\begin{align}
\Omega_{bm}=1-\left(\frac{\eta_{RG}}{\nu_{RG}}\right) \frac{\left(1+\omega_{ki}\right)}{\left(\omega_{bm}-\omega_{ki}\right)}~.
\label{sacds}
\end{align} 
At that fixed point, we obtain the evolution equations of $x$ and $y$ as 
\begin{align}
\frac{\mathrm{d}x}{\mathrm{d}N} &=\frac{3 y^{2}}{2 \left(1-y^{2}\right)} \left[ \sigma \left( 1-x^{2}\frac{\epsilon_{RG}}{\nu_{RG}}-y^{2} \right)-\frac{\left(1+\omega_{bm}\right)}{x} \left( 1-x^{2}-y^{2} \right) \right]
\label{s2fixx} \\ \frac{\mathrm{d}y}{\mathrm{d}N}& = -\frac{3}{2}\sigma x y (1-\frac{\epsilon_{RG}}{\nu_{RG}})~.
\label{s2fixy}
\end{align}
The evolution equations ensure that for the existence of this scaling fixed point, two conditions needed to be fixed as $\epsilon_{RG}=\nu_{RG}$ and $\sigma_{*}=\frac{\left(1+\omega_{bm}\right)}{x_{*}}$ which will give $\frac{\mathrm{d}x}{\mathrm{d}N}=0$ and $\frac{\mathrm{d}y}{\mathrm{d}N}=0$. Moreover, we have obtained the RG scale evolution for the scaling fixed point, putting first the fixed point value of $\sigma_{*}$ at Eq. \eqref{fllnk}
\begin{align}
\frac{\mathrm{d}\ln k}{\mathrm{d}N}=-\frac{2}{\eta_{RG}+\nu_{RG}} \left[
\frac{3}{2} \frac{\epsilon_{RG}}{ \nu_{RG}}\left(1+\omega_{bm}\right)-\frac{3}{2}\left(1+\omega_{bm}\right)\right]~.
\label{slnk}
\end{align}
Now for $\epsilon_{RG}=\nu_{RG}$, the RG scale freezes at some scale $k_{ki}$, which confirms that this scaling fixed point only exists when $\frac{\mathrm{d}\ln k}{\mathrm{d}N}=0$. Setting $\frac{\mathrm{d}\sigma}{\mathrm{d}N}=0$ from Eq. \eqref{chsigma} and substituting the value of $\sigma_{*}$ into this equation, we obtain $\Gamma$ as a constant given by the following expression:
\begin{align}
\Gamma-1=\frac{\left(1-\omega_{ki}\right)}{\left(1+\omega_{ki}\right)}+\frac{\left(1-\omega_{ki}\right)\left(1-\omega_{bm}\right)}{\left(1-\omega_{ki}\right)\left(\omega_{bm}-\omega_{ki}\right)}~.
\label{gammaf}
\end{align}
For the canonical case where $\omega_{ki}=1$, we end up with exponential potential as $\Gamma=1$ but for the non-canonical case where $\omega_{ki} \neq 1$, we obtain a power law potential profile as $\Gamma$ becomes constant involving the non-canonical parameter $\omega_{ki}$ and the equation of state of background fluid $\omega_{bm}$.  
%
%
\section{Conclusions}
In this paper, we have explored the dynamical equations in the presence of a non-canonical scalar field and a background matter, taking into account quantum corrections using the functional renormalization group. The important point we want to address here is how the non-canonical scalar field parameter influences the dynamical equations; this in turn affects the cosmic evolution of the universe when quantum modifications are included. Here, we have particularly studied cosmic evolution by considering the running of the gravitational constant and the non-canonical scalar field potential. We have considered the covariant conservation of energy-momentum tensor by both the non-canonical scalar field and the background matter separately. This in turn gives another constraint  \eqref{flrg} in order to preserve the Bianchi identity. This constraint relation between the RG flow parameters and the dimensionless cosmological variables has particular importance because it restricts the arbitrary behaviour of the RG flow in Einstein-Hilbert truncation.  \\
In this work, we study the dynamical equations showing the difference from the canonical case and how this evolution depends on the non-canonical scale parameter $\omega_{ki}$, and the RG flow parameters, namely, $\eta_{RG}$, $\nu_{RG}$ and $\epsilon_{RG}$. The  equation of state for the kinetic energy of non-canonical scalar field $\omega_{ki}$ represents how this evolution deviates from the canonical case where $\omega_{ki}=1$; which signifies $F(X)=X$. We have identified various fixed points, namely, the potential-dominated fixed points, the kinetic energy dominated fixed points, the mixed fixed points, and the scaling fixed points depending on the energy contributions from the non-canonical scalar field and the background matter fields. The cosmic fixed points have been studied for a Lagrangian in the form of $\mathcal{L}(\phi, X)=F(X)-V_{k}(\phi)$ with a kinetic term given by $F(x)=AX^{n}$ and with a power law or exponential potential. In this study, the cosmological fixed points are attained only when the RG parameters become constant or $k$-independent. This condition exclusively involves RG fixed points, resulting in a constant evolution of the RG scale. \\
The cosmic fixed points exclusively depend on the non-canonical scalar field parameter $\omega_{ki}$ though the constraint \eqref{flrg} involving the RG parameters and dimensionless cosmological quantities. The cosmic fixed points which we identified from dynamical equations may give an accelerated universe depending on the value of the effective equation of state $\omega_{eff}$ and the RG parameters. For the mixed fixed points, we observed the possibility of achieving a dark energy-dominated accelerated universe, depending on the non-canonical equation of state of kinetic energy $\omega_{ki}$ for different RG theories. An important result of our work is that we have obtained a bound on the ratio of the RG parameters $\frac{\eta_{RG}}{\nu_{RG}}$ involving this non-canonical parameter to attain an accelerated ratio of RG parameters universe for the mixed fixed points. This bound plays a significant role as the non-canonical scalar field includes a larger set of RG fixed points, which may lead to an accelerated universe.
%
For example, at the asymptotically safe matter RG fixed point in the UV scale, where $\eta_{RG} \rightarrow -2$ and $\nu_{RG} \rightarrow 4$, this universe does not transition to an accelerated state for this fixed point in the canonical case where $\omega_{ki}=1$. However, surprisingly, for the non-canonical case, for the same values of RG parameters, it results in an accelerated universe at an early epoch for $\omega_{ki}<\frac{1}{3}$. This, in turn, imposes a bound on the exponent of the kinetic term as $n>2$.
Moreover, the potential-dominated cosmic fixed point could potentially pave the way for an accelerated universe. This universe might be analogous to a dark energy-dominated or inflating universe, depending upon the diverse range of possible values for the RG parameters within the specific renormalization group theory.
In the kinetic-dominated cosmic fixed point, a noteworthy distinction emerges: unlike the canonical case where only one cosmic fixed point emerges in the near infrared region, here, two cosmic fixed points coexist. These fixed points are manifested when the gravitational coupling reaches the Gaussian RG fixed points and becomes constant. However, this particular fixed point does not inherently lead to a dark energy-dominated or inflating universe unless the exponent of the non-canonical kinetic term is less than $-1 $, which would be considered highly unlikely.
%
%
Another essential cosmic fixed point is the scaling fixed point, where the energy density of the non-canonical scalar field scales with the energy density of background matter. For the first scaling point denoted by d1 in table \eqref{table-I}, we have seen that it exactly matches with the canonical case and the cut off scale $k$ is directly proportional to the Hubble parameter $H$.  In the second scaling fixed point (d2), we illustrate how these fixed points vary depending on the non-canonical scalar parameter when quantum corrections are included. In this scenario, the potential profile can be precisely determined based on whether the scalar field is canonical or non-canonical.
%
\section*{Acknowledgment} RM would like to thank IOE-UOH for financial support through the IOE-PDRF scheme.


\begin{thebibliography}{99}
%
\bibitem{Riess:1998}
A. G. Riess et al., ``Observational evidence from supernovae for an accelerating
universe and a cosmological constant",
Astron. J. {\bf 116}, 1009 (1998).
%
\bibitem{Liddle:2000}
A. R. Liddle and D. Lyth, ``Cosmological Inflation and Large Scale Structure" (Cambridge University Press, Cambridge, England, 2000)
%
\bibitem{Tsujikawa:2006}
E. J. Copeland, M. Sami, and S. Tsujikawa, ``Dynamics of dark energy",  Int. J. Mod.
Phys. D {\bf 15}, 1753 (2006).
%
\bibitem{Copeland:1998}
E. J. Copeland, A. R. Liddle, D. Wands, ``Exponential potentials and cosmological scaling solutions", Phys. Rev. D {\bf 57}, 4686-4690 (1998).
%
\bibitem{Wilson:1971}
K.G. Wilson, ``Renormalization group and critical phenomena. I. Renormalization group and
the Kadanoff scaling picture". Phys. Rev. B {\bf 4}, 3174–3183 (1971).
%
\bibitem{Reuter:1996cp}
M.~Reuter,
``Nonperturbative evolution equation for quantum gravity,''
Phys.\ Rev. \ D {\bf 57}, no. 10, 971 (1998).
%
\bibitem{Souma:1999at} 
W.~Souma,
``Nontrivial ultraviolet fixed point in quantum gravity,''
Prog.\ Theor.\ Phys.\  {\bf 102}, 181 (1999).
%
\bibitem{Lauscher:2001ya} 
O.~Lauscher and M.~Reuter,
``Ultraviolet fixed point and generalized flow equation of quantum gravity,''
Phys.\ Rev.\ D {\bf 65}, 025013 (2002).
%
\bibitem{Litim:2004}
D.F. Litim, ``Fixed points of quantum gravity", Phys. Rev. Lett. {\bf 92}(20), 201301 (2004).
%
\bibitem{Niedermaier:2006wt} 
M.~Niedermaier and M.~Reuter,
``The Asymptotic Safety Scenario in Quantum Gravity,''
Living Rev.\ Rel.\  {\bf 9}, 5 (2006).
%
\bibitem{Weinberg:1976}
S. Weinberg, ``Critical phenomena for field theorists", in Proceedings 14th International School
of Subnuclear Physics, Erice (1976), p. 1.
%
%

\bibitem{Bonanno:2001xi}
A.~Bonanno, M.~Reuter,
``Cosmology of the Planck era from a renormalization group for quantum gravity",
Phys.\ Rev.\ D {\bf 65}, 043508 (2002).
%
\bibitem{Koch:2011}
B. Koch and I. Ramirez, ``Exact renormalization group with optimal scale and its application to
cosmology", Class. Quant. Grav. {\bf 28} 055008 (2011).
%
\bibitem{Weinberg:2010}
S. Weinberg, ``Asymptotically safe inflation", Phys. Rev. D {\bf 81} 083535 (2010).
%
\bibitem{Contillo:2011}
A. Bonanno, A. Contillo and R. Percacci, ``Inflationary solutions in asymptotically safe $f(R)$
theories", Class. Quant. Grav. {\bf 28} 145026 (2011).
%
\bibitem{Bonanno:2002}
A. Bonanno and M. Reuter, ``Cosmology with self-adjusting vacuum energy density from a
renormalization group fixed point", Phys. Lett. B {\bf 527} 9 (2002).
%
\bibitem{Benti:2004}
E. Bentivegna, A. Bonanno and M. Reuter, ``Confronting the IR fixed point cosmology with high
redshift supernova data", JCAP {\bf 01} 001 (2004).
%
\bibitem{Shapiro:2005}
I.L. Shapiro, J. Solà and H. Stefancic, ``Running G and Lambda at low energies from physics at
$M(X)$: possible cosmological and astrophysical implications", JCAP {\bf 01} 012 (2005).
%
\bibitem{Mandal:2020}
R. Mandal, S. Gangopadhyay and A. Lahiri , ``Cosmology of Bianchi type-I metric using renormalization group approach for quantum gravity", Class. Quant. Grav. {\bf 37} (2020) 065012.

%
\bibitem{Hindmarsh:2011}
M. Hindmarsh, D. Litim and C. Rahmede, ``Asymptotically safe cosmology", JCAP {\bf 07}, 019 (2011).
%
\bibitem{Linder:2011}
C. Ahn, C. Kim and E. V. Linder, ``From asymptotic safety to dark energy", Phys. Lett. B {\bf 704}, 10–14 (2011).
%
\bibitem{Sen:2002}
A. Sen, Mod. Phys. Lett. A {\bf 17}, 1797 (2002);


\bibitem{Sachs:2003}
N. D. Lambert and I. Sachs, Phys. Rev. D {\bf 67}, 026005 (2003).


\bibitem{Mukhanov:1999}
C. Armendariz-Picon, T. Damour, and V. F. Mukhanov,  ``k-Inflation", Phys. Lett. B
{\bf 458}, 209 (1999).
%
\bibitem{Mukhanov:1999a}
J. Garriga and V. F. Mukhanov, ``Perturbations in $k$-inflation", Phys. Lett. B 458, 219
(1999).
%
\bibitem{Unnikrishnan:2012}
S. Unnikrishnan, V. Sahni and A. Toporensky, ``Refining inflation using non-canonical scalars" J. Cosmol. Astropart. Phys. {\bf 1208}, 018
(2012). 

\bibitem{Chiba:2000}
T. Chiba, T. Okabe, and M. Yamaguchi, 
``Kinetically driven quintessence," 
Phys. Rev. D {\bf 62}, 023511, (2000).

\bibitem{Steinhardt:2000}
C. Armendariz-Picon, V. F. Mukhanov, and P. J. Steinhardt, ``A dynamical solution
to the problem of a small cosmological constant and late-time cosmic acceleration,"
Phys. Rev. Lett. {\bf 85}, 4438 (2000).
%
\bibitem{Steinhardt:2001}
C. Armendariz-Picon, V. F. Mukhanov, and P. J. Steinhardt, ``Essentials of
k-essence," Phys. Rev. D {\bf 63}, 103510 (2001).
%
\bibitem{Chiba:2002}
T. Chiba, ``Tracking $k$-essence", Phys. Rev. D {\bf 66}, 063514 (2002).
%
\bibitem{Scherrer:2004}
R. J. Scherrer, ``Purely Kinetic $k$ Essence as Unified Dark Matter", Phys. Rev. Lett. Phys. Rev. Lett. 93, 011301 (2004).
%
\bibitem{Chimento:2004}
L. P. Chimento, ``Extended tachyon field, Chaplygin gas, and solvable $k$-essence cosmologies" Phys. Rev. D 69, 123517 (2004).
%
\bibitem{Bose:2009}
N. Bose, A. Majumdar, ``A $k$-essence model of inflation, dark matter
and dark energy" Phys. Rev. D {\bf 79}, 103517 (2009).
%
\bibitem{Wands:2013}
Josue De-Santiago, Jorge L. Cervantes-Cota and David Wands, ``Cosmological phase space analysis of the $F(X)- V(\phi)$ scalar field and bouncing solutions", Phys. Rev. D {\bf 87}, 023502 (2013).
%
\bibitem{Cota:2011}
J.De-Santiago, J. L. Cervantes-Cota, ``Generalizing a unified model of dark matter, dark energy, and inflation with non-canonical kinetic term", Phys. Rev. D {\bf 83}, 063502 (2011). 
%
\bibitem{Sahni:2017}
Varun Sahni, Anjan A. Sen, ``A new recipe for $\Lambda$CDM", Eur. Phys. J. C {\bf 77}, 225 (2017).
%
\bibitem{Sahni:2021}
S. S. Mishra, V. Sahni, ``Unifying dark matter and dark energy with non-canonical scalars", Eur. Phys. J. C {\bf 81}, 625 (2021).
%
\bibitem{Fang:2007}
Wei Fang, H Q Lu and Z G Huang
``Cosmologies with a general non-canonical scalar field", Class. Quantum Grav. {\bf 24}, 3799–3811 (2007).
%
\bibitem{Reuter:2001ag} 
M.~Reuter and F.~Saueressig,
``Renormalization group flow of quantum gravity in the Einstein-Hilbert truncation,''
Phys.\ Rev.\ D {\bf 65}, 065016 (2002)
%
\bibitem{Percacci:2003}
R. Percacci and D. Perini, ``Asymptotic safety of gravity coupled to matter", Phys. Rev. D {\bf 68} 044018 (2003).
%
\bibitem{Narain:2010}
G. Narain and R. Percacci, ``Renormalization group flow in scalar-tensor theories. I,
Class. Quant. Grav. {\bf 27}, 075001 (2010).
%
\bibitem{Rahmede:2010}
G. Narain and C. Rahmede, Renormalization group flow in scalar-tensor theories. II,
Class. Quant. Grav. {\bf 27}, 075002 (2010).







\end{thebibliography}
\end{document}